# Seeking for low thermal conductivity atomic configurations in $Si_{0.5}Ge_{0.5}$ alloys with Bayesian Optimization

**Jiahao Yan,**[a#] **Han Wei,**[a#] **Han Xie,**[a, b] **Xiaokun Gu** [c] **and Hua Bao** [a*]

[a] *University of Michigan-Shanghai Jiao Tong University Joint Institute, Shanghai Jiao Tong University, Shanghai 200240, China*

[b] *Artificial Intelligence Lab, SAIC Motor Corporation, Shanghai 200433, China*

[c] *Institute of Engineering Thermophysics, School of Mechanical Engineering, Shanghai Jiao Tong University, Shanghai 200240, China*

---

\* Corresponding Author

E-mail: hua.bao@sjtu.edu.cn. #These authors contributed equally

**Abstract:** The emergence of data-driven science has opened up new avenues for understanding the thermophysical properties of materials. For decades, alloys are known to possess very low thermal conductivity, but the extreme thermal conductivity can be achieved by alloying has never been identified. In this work, we combine the Bayesian optimization with a high throughput thermal conductivity calculation to search for the lowest thermal conductivity atomic configuration of $Si_{0.5}Ge_{0.5}$ alloy. It is found layered structures are most beneficial for reducing the thermal conductivity among all atomic configurations, which is attributed to the strong branch-folding effect. Furthermore, the roles of interface roughness and layer thicknesses in producing the lowest thermal conductivity are investigated. Through another comprehensive search using Bayesian optimization, the layered structure with smooth interfaces and optimized layer thickness arrangement is identified as the optimal structure with the lowest thermal conductivity.

# 1. Introduction

In the field of thermal science and material science, machine learning has been regarded as the "fourth paradigm", which provides new approaches to develop understandings in thermophysical characteristics of materials [1-5]. Machine learning algorithms have been successfully implemented to extract the structural features and establish the structure-property linkage [6-9]. Owing to the advantages of efficient global search, machine learning algorithms also have benefits in structure optimization and design [10,11]. For example, optimization algorithms, such as genetic algorithm, Bayesian optimization algorithm, have been proven to be very effective in optimizing structures of 1D random multilayer [12], 2D nanoporous graphene [13,14] and 3D composite interface[15], etc. Here, we consider applying machine learning algorithm for optimizing the atomic configuration of alloy structures.

Lattice thermal conductivity is one of the most important material thermophysical properties[16]. Materials with low thermal conductivity have many applications in thermal insulation, thermoelectric energy conversion, and thermal management in microelectronics devices [17-20], etc. Careful design of the atomic configuration plays an important role in manipulating the thermal conductivity. Alloying has been proposed as an effective method to reduce thermal conductivity due to the mass and bond difference [21-25]. Sharp decrease of lattice thermal conductivity has been observed after a small amount of alloying [21]. In well-developed disordered alloys, the phonon lifetimes can be minimized to the order of the time scale of thermal vibrations [25,26]. For a certain composition, the thermal conductivity of disordered alloy is often considered as the lower bound of thermal conductivity of any material, which is called the alloy limit [27,28]. Recently, many efforts have been endeavored to break the alloy limit by constructing phononic crystals [17,29-33]. The phonon propagation is greatly blocked by the periodic or aperiodic scattering inclusions in phononic crystals, leading to the localization of certain phonon modes [34,35]. Although ultra-low thermal conductivity can be achieved through designing the phononic crystals under the guidance of physical intuition, previous conclusions were based only on the study of limited structures. For a random composite, the characteristics of atomic configuration with low thermal conductivity have not been systematically investigated. A comprehensive search for the lowest thermal conductivity structure, to the best of our knowledge, is still lacking.

Numerical simulations can be applied to investigate the thermal transport problems, including molecular dynamics and first-principles-based approaches. Due to the relatively large computational cost[36-38], if an inefficient manual search among a large number of candidates is adopted, the total computational cost will be unaffordable. Therefore, machine-learning optimization algorithms are more preferable. In this work, we consider $Si_{0.5}Ge_{0.5}$ alloy system and search for the optimal atomic configuration with the lowest thermal conductivity using Bayesian optimization. First-principles-based approaches are adopted to compute the thermal conductivities of alloys with different atomic configurations. We first investigate the general characteristics of the low thermal conductivity structure. Then, spectrum analyses of the phonon transport properties of typical atomic configurations are performed to reveal physics behind the low thermal conductivity. Finally, the specific atomic configuration with the lowest thermal conductivity is obtained by optimizing the detailed atom arrangement.

## 2. Methodology

We search for the optimal atomic configuration of $Si_{0.5}Ge_{0.5}$ alloy with the Bayesian optimization-based framework. The optimization unit cell is composed of half Si atoms (yellow spheres) and half Ge atoms (red spheres), as shown in Fig. 1. Since the lattice constants of Si and Ge are similar in magnitude, the lattice mismatch between Si and Ge is ignored and the lattice constants are set as the same as Si [15]. The size of the unit cell is characterized by the number of face-centered-cubic (FCC) conventional unit cells along the $x$, $y$ and $z$ directions. For instance, a unit cell of size 4×2×2 denotes 4, 2, and 2 FCC conventional cells along the $x$, $y$, and $z$-direction, respectively. The different atomic configurations of alloys are generated by randomly assigning the atomic masses at the lattice points as Si or Ge while maintaining the composition. We note some configurations are symmetric from the thermal transport point of view. Thus, after constructing different atomic configurations, we remove these symmetric configurations according to their translational and rotational invariances [39,40]. The remained configurations consist of the whole set of candidates for optimization.

Bayesian optimization is a machine-learning-based optimization method, which is well suited for the optimization of "black-box" objective functions that have no explicit expressions and are computationally expensive to evaluate[41,42]. Recently, Bayesian optimization has become very popular in automatic machine learning, robotics, experiment and material design [43-45], etc. Therefore, in this study, we apply Bayesian optimization to search for the optimal atomic configurations. The optimization workflow is shown in Fig. 1. First, initial candidate configurations (20, in this work) are randomly selected from the whole set of candidates and the corresponding lattice thermal conductivities are computed. Through Bayesian optimization, the optimized structure among the candidates is selected according to specific strategies (see Supporting Information). The optimization process will iterate continuously until the convergence is reached and then the optimal configuration is output. The convergence criterion is that 80 percent of thermal conductivities of calculated candidate configurations have no changes comparing to the last iteration. Open-source Bayesian optimization library COMBO is employed to optimize the atomic configurations [46]. More details about Bayesian optimization are included in the Supporting Information. For all the configurations, the lattice thermal conductivities are along the $x$-direction, which are obtained by first-principle based calculations (see more details in Supporting Information).

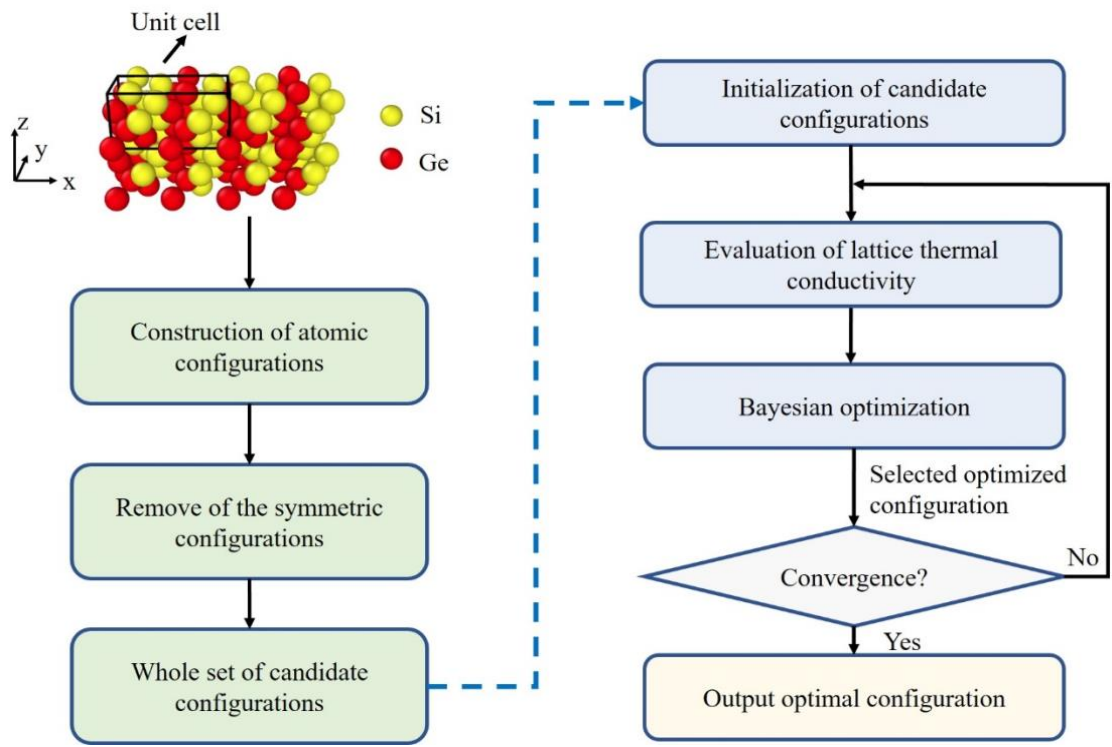

**Fig. 1** Schematic of the Bayesian optimization-based framework for the optimization of atomic configurations. The black box shows the unit cell, in which the yellow and red spheres denote Si and Ge atoms, respectively. The different atomic configurations are generated by randomly assigning the atoms at lattice points as Si or Ge while maintaining the composition of Si and Ge both as 0.5. Symmetric configurations are further filtered according to their translation and rotation invariance. The remained configurations are composed of the whole set of candidates. First-principles based methods are implemented to evaluate the lattice thermal conductivities. Through Bayesian optimization, the optimal configuration can be selected after a few iterations.

## 3. Structure with low thermal conductivity

We start with the searching of low thermal conductivity atomic configurations in unit cells of size 2×1×1, 2×2×1, and 2×2×2. For 2×1×1 unit cell. Since the whole set of candidates only consist of 652 configurations, we simply compute the thermal conductivities of all candidates without further optimization. The structure with the lowest thermal conductivity is shown in Fig. 2(a), where silicon and germanium atoms are gathered together, forming a layered structure. Fig. 2(b) shows the structure with the highest thermal conductivity, which has alternating Si/Ge monoatomic layers.

Note that the computational cost sharply increases with increasing unit cell size (2×2×2 unit cell possesses over $10_{18}$ candidates). In order to reduce the computational cost, we group neighboring atoms as new units to reduce the degree of freedom of systems. After grouping the atoms, the number of candidates for 2×2×1, and 2×2×2 unit cells are reduced to 963 and 453, respectively. More details about the strategy of grouping neighboring atoms can be seen in the Supporting Information. Similar to the 2×1×1 unit cell, we compute the thermal conductivities of all configurations. The optimal configurations with the lowest thermal conductivity of 2×2×1 and 2×2×2 unit cells are shown in Fig. 2(c) and (e), respectively. They have similar characteristics to the configuration in Fig. 2(a) that the Si and Ge atoms are aggregated to form layered structures. The corresponding configurations with the highest thermal conductivity are shown in Fig. 2(d) and (f), which have alternating Si/Ge monoatomic layers similar to the configuration in Fig. 2(b).

With the full calculation of all the candidates of three different unit cell sizes, we find that layered structures have lower thermal conductivities. To confirm this, the alloy system with a large unit cell of size 6×1×1 is studied. Since the number of candidates is huge (about $10^{14}$), the computations of all the candidates are not feasible. Therefore, Bayesian optimization is performed to search for the optimal configurations (the workflow is shown in Fig. 1). Note that the full anharmonic calculation is too time-consuming for 6×1×1 unit cell. To reduce the computational cost, a relaxation time model proposed by Klemens[47] is adopted to calculate the phonon relaxation time, while the phonon group velocity and specific heat are still obtained from the first-principles calculations. We have demonstrated that using the simplified model to calculate thermal conductivities does not induce significant deviation from the full calculation using the 2×1×1 cell. More details can be seen in the Supporting Information. In spite of this, Bayesian optimization still cannot obtain converged results due to the huge amount of candidates. Nevertheless, the structural characteristics with low thermal conductivity can still be inferred by the optimization. We use the layered coefficient to quantify the layered state of configurations, of which the schematic is shown in Fig. 2 (g). First, a small box moving one monolayer at one step scans through the configuration along the *x*-direction, and it is defined that if all atoms in the box are of the same type, these atoms form a layer. The total number of layers recorded by the moving box is defined as the layered coefficient. A larger layered coefficient denotes the atomic configuration is closer to a layered structure. Note that a too thick scanning box will cause the omission of thin layers. A too thin scanning box will lead to large numbers of layers in most structures so that the layered states cannot be distinguished. After balancing those factors, the thickness of the scanning box is set to three monatomic layers.

Three independent optimizations with different initial candidate datasets are carried out. The evolutions of thermal conductivities and the number of layers of optimized configurations are shown in Fig. 2(i). As the optimization proceeds, the thermal conductivities of optimized structures gradually decrease. The optimized thermal conductivities are about 30% lower than that of the initial configurations. Overall, the numbers of layers increase as the optimizations progress, indicating that the optimized configuration with lower thermal conductivity gradually approaches to layered structures. The configurations with the lowest and highest conductivity among the searched candidates are shown in Fig. 2(g) and Fig. 2(h), respectively, and are also marked in Fig. 2(i). Apparently, the low thermal conductivity configuration (Fig. 2(g)) has a more distinct layered characteristic compared with the high thermal conductivity one (Fig. 2(h)), which is consistent with Fig. 2 (a)-(f).

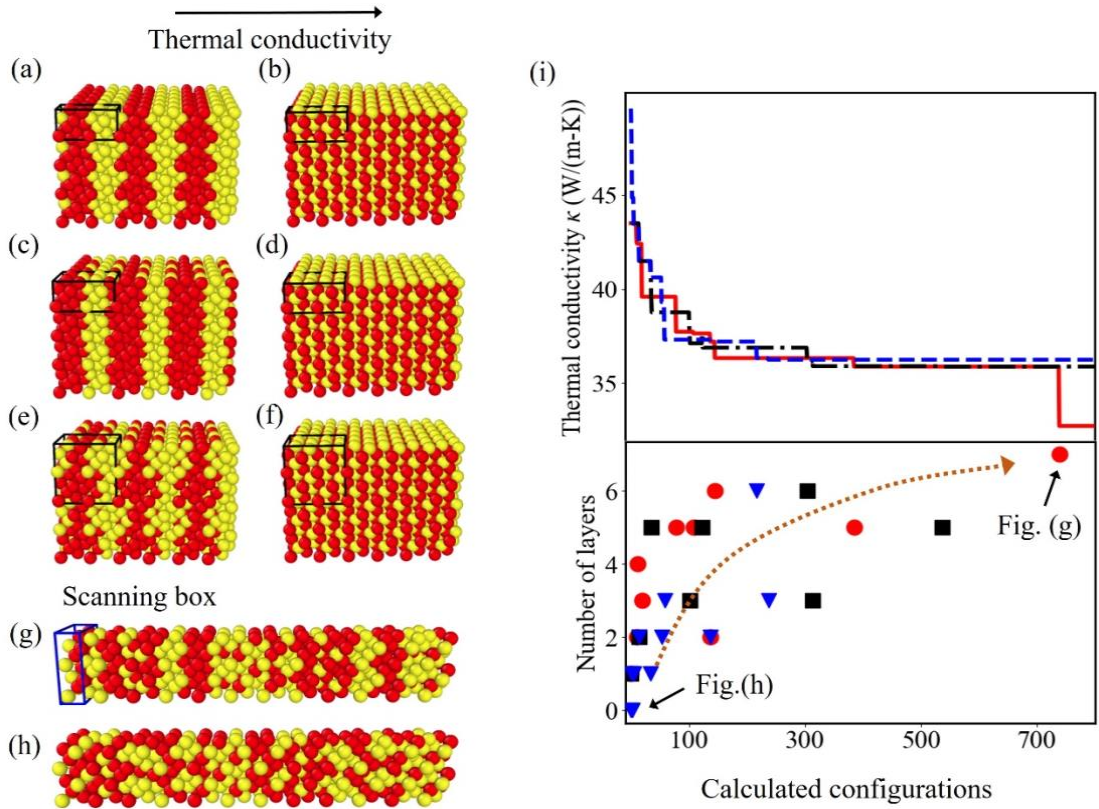

**Fig. 2** The search for low thermal conductivity atomic configurations of different unit cell sizes. The lowest thermal conductivity configuration of (a) 2×1×1, (c) 2×2×1, and (e) 2×2×2 unit cells. The highest thermal conductivity configuration of (b) 2×1×1, (d) 2×2×1, and (f) 2×2×2 unit cells. The configuration with the highest (g) and lowest conductivity (h) among the searched candidates in Bayesian optimization of 6×1×1 unit cell, respectively. (i) The evolutions of thermal conductivities and layered coefficients of optimized configurations of 6×1×1 unit cell. The red, blue, and black lines (dots) denote three independent optimizations.

We further analyze the cause of the low thermal conductivity in layered structures. It is noted that the same type of atoms in the lowest and highest thermal conductivity cases are aggregated (as shown in Fig. 2 a, c, e, g) and dispersed (as shown in Fig. 2 b, d, f, h), respectively. Therefore, we investigate the relationship between thermal conductivities and the state of aggregation of the same kind of atoms, which can be quantified by short range order coefficient (*SROC*)[48]. The *SROC* is defined as

$$SROC = 1 - \frac{P}{M_A},  \qquad (1)$$

where *P* denotes the probability of finding the atom *A* at nearest neighbor positions of the atom *B*, and $M_A$ is the atom fraction of *A* atoms (equal to 0.5 in this work). The atomic configurations with aggregated and dispersed atoms of the same kind are clustering and short range order structures, and the corresponding values of *SROC* are positive and negative, respectively. Other atomic configurations with *SROC* equal to zero are classified as disordered crystals.

We take the configurations of the 2×1×1 unit cell as an illustration. All the 652 configurations are divided into clustering structures, short range order structures and disordered crystals according to the *SROC* coefficients. The distribution of thermal conductivities of these configurations is shown in Fig. 3. Generally, the thermal conductivities of short range ordered structures are higher than those of disordered crystals, followed by the clustering structures, indicating that a more aggregated state of atoms can lead to lower thermal conductivity.

Usually, the aggregation state of atoms can affect the number of atoms in the primitive unit cell, which is strongly correlated with the lattice thermal conductivity [49-51]. We further examine the number of atoms in primitive unit cells of different configurations, which are also presented in Fig. 3. It is seen that the lowest and highest thermal conductivity configurations have the largest (16) and smallest (2) number of atoms in the primitive unit cell, respectively. From the perspective of phonon transport, large number of atoms in primitive unit cell can induce the shrink of Brillouin zone[49,52-54], shown as the folding and flattening of phonon dispersion curve, which can reduce the thermal conductivity (will be discussed in detail later). On the contrary, a simple configuration, with few atoms in the primitive unit cell can have a high lattice thermal conductivity. Note that there are some clustering structures with abnormal high thermal conductivities, which are circled by a dotted line in Fig. 3. One reason for the high thermal conductivity is the small number of atoms in the primitive unit cell. Another is the formation of the transport path for coherent phonons constructed by the layers paralleling to the direction of thermal conductivities [15,27,55], as shown in the inset figure in Fig. 3.

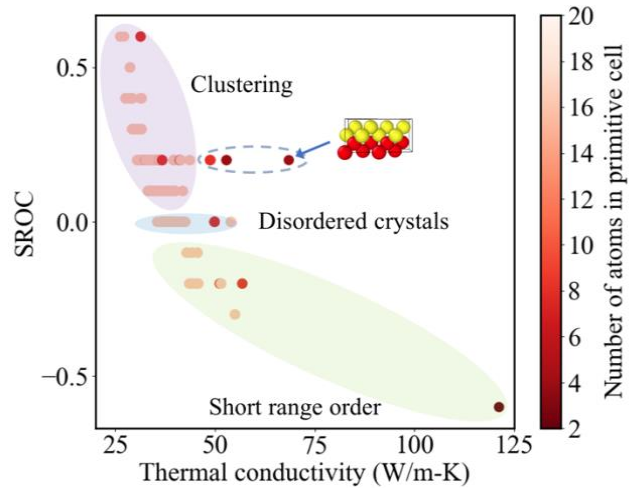

**Fig. 3** The thermal conductivities, short range order coefficient (SROC) and numbers of atoms in the primitive unit cell of atomic configurations of 2×1×1 unit cell. For clustering structures, disordered crystals, and short range order structures, the SROCs are positive, zero and negative, respectively.

In addition to the aggregate state and number of atoms in the primitive unit cell, we investigate the phonon transport characteristics of low thermal conductivity. Three typical configurations of 2×1×1 unit cell are analyzed: the layered structure with the lowest thermal conductivity (shown in Fig. 2(a)), a disordered crystal with the intermediate thermal conductivity, and the structure with alternating Si/Ge monoatomic layers (alternating monolayers) with the highest thermal conductivity (shown in Fig. 2(b)). Spectrum analyses of thermal transport properties are performed, including dispersion curve, group velocity $v$, relaxation time $\tau$, the density of state (DOS) and the DOS weighted thermal conductivity $\langle \kappa(w) \rangle$. For simplicity, the implementation of spectrum analysis is not included here, and more details can be found in references [21,55].

The DOS weighted thermal conductivities $\langle \kappa(w) \rangle$ of the layered structure, disordered crystal, and alternating monolayers structure are shown in Fig. 4(a). It can be seen that the thermal conductivities of the layered structure and the disordered crystal are dominated by the low-frequency phonons, around 0-5 THz. However, for the alternating monolayers, the phonons with frequencies higher than 10 THz have large contributions to the thermal transport.

The dispersion curves of the layered structure and disordered crystal within the frequency range of 0-5 THz are shown in Fig. 4(c). It is seen that compared to the disordered crystal (gray lines), the dispersion curve of the layered structure (red lines) is more flattened. Previous studies have reported that the flattening of the dispersion curve can result in the reduction of phonon group velocity [52,54]. The square of phonon group velocities of three configurations are shown in Fig. 4(b). It is seen that in general, the layered structure has lower phonon group velocities, which is consistent with the results in Fig. 4(c). On the other hand, the flattening of the phonon dispersion curve suggests that more phonon modes are forbidden and thus the thermal transport is weakened.

The DOS of three configurations and Si, Ge are depicted in Fig. 4(d). Overall, DOS of the layered structure is lower in the diverged frequency range of Si and Ge comparing to the other two structures. In other words, there are fewer available phonon modes for heat conduction in this range, which is consistent with the implications of Fig. 4(c). In addition to the filtering effect of alternate Si/Ge layers, the reflection and interference of phonons between layers can also forbid certain phonon states. From the spectrum analysis, it is reasonable that layered structure can have low thermal conductivity.

The high thermal conductivity of alternating monolayers can also be deduced from the spectrum analyses. The dispersion curves of the alternating monolayers structure and disordered crystal are shown in Fig. 4(f). It is seen that the alternating monolayers structure has a large bandgap and a narrow optical energy band, which can lead to a reduction of scattering between acoustic phonons and optical phonons. Therefore, the phonon lifetimes in the corresponding frequency ranges will be large. The phonon relaxation time of the three configurations is shown in Fig. 4(e). Overall, the alternating monolayers structure has larger phonon relaxation time than that of the other two structures, which is consistent with Fig. 4 (f). Note that structure with alternating thin layers has been adopted to enhance the thermal conductivity in many materials [51,54,56,57]. This study can provide evidence of the optimality of alternating monolayers for the highest thermal conductivity in the alloy systems.

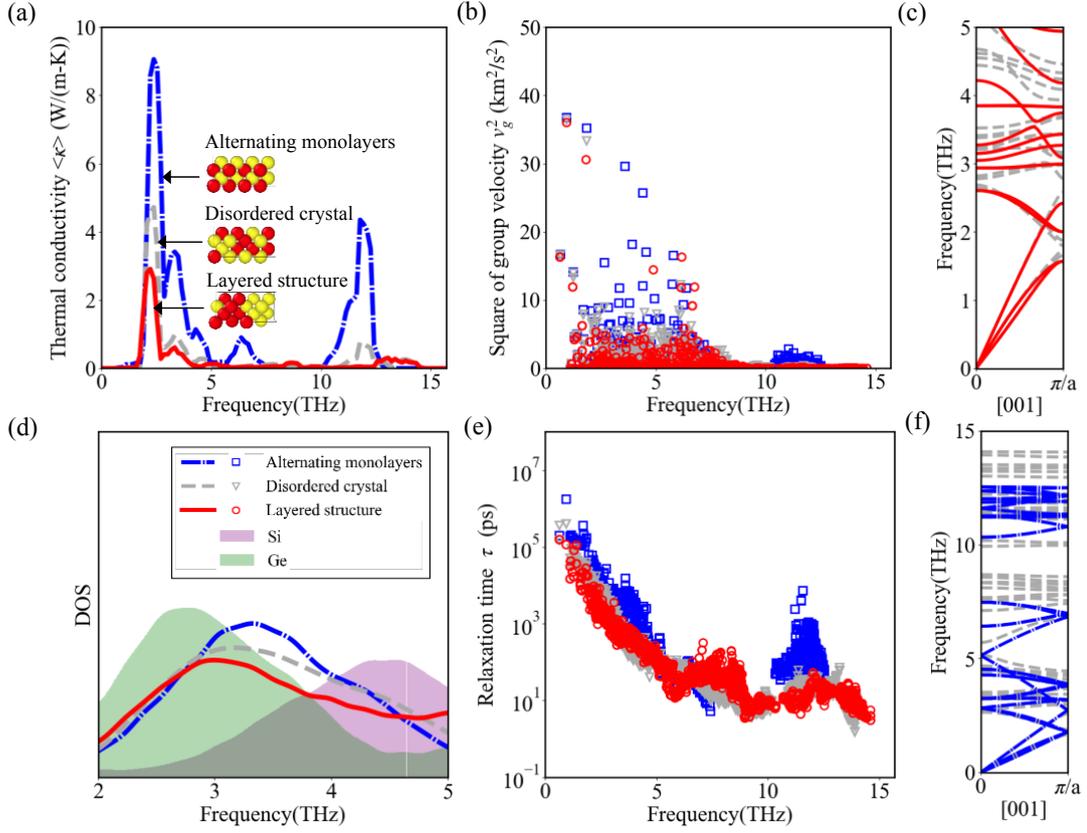

**Fig. 4** Spectrum analyses of thermal transport properties of the layered structure, the disordered crystal, and the alternating monolayers structure. (a) DOS weighted thermal conductivity. (b) Phonon group velocity. (c) Comparison of dispersion curves between the disordered crystal (gray lines) and the layered structure (red lines). (d) DOS. (e) Phonon relaxation time. (f) Comparison of the dispersion curves between the alternating monolayers structure (blue lines) and the disordered crystal (gray lines). In (c) and (f), a is the lattice constant of the 2×1×1 unit cell in [001] direction.

## 4. Atomic configurations of the layered structure with the lowest thermal conductivity

From the investigations above, the atomic configuration with low thermal conductivity is identified as the layered structure. Previous studies have shown that the interface roughness [28,58-60] and different thicknesses of alternating layers [27,33,61] can significantly affect the thermal conductivity of layered structures. In this section, we further search for the specific configuration of layered structure with the lowest thermal conductivity by manipulating the interface roughness and arrangement of alternating layer thickness. Since it is not easy to consider the two factors at the same time, we first consider one factor at a time.

**4.1 Optimizations of interface roughness and layer thickness arrangement separately**

First, the optimization of interface roughness for configurations of 4 × 2 × 1 unit cells is carried out. Two unit cells along the $z$-direction are set to provide a large design degree of freedom for optimization the interface. For simplicity, we set the Si/Ge layers with the same thickness. The schematic of the optimization system is shown in Fig. 5 (a). Along $x$-direction (from left to right), the unit cell contains, half of Si layer, interface I, the Ge layer, interface II, and the other half of the Si layer. In order to generate different configurations, the atoms in interface I are randomly assigned as Si or Ge. Interface II is then constructed by exchanging Si and Ge atoms corresponding to the interface I to maintain the composition. There are 928 candidates in total, and Bayesian optimization is performed to search for the optimal configuration. The evolution of optimized thermal conductivities is similar to Fig. 2 (i).

In this scenario, after seven drops of the searched lowest thermal conductivity, the converged solution is obtained. All the calculated configurations can be divided into 8 groups according to the evolution of the optimization process. The highest and lowest thermal conductivity configurations exist in group 1 and 8, respectively. In order to investigate the characteristics of the optimized interfaces, three sets of configurations from groups 1, 4 and 8 are taken as representatives, of which the schematics are shown in Fig. 5(c). From group 1 to group 8, the extent of mixing of two kinds of atoms at the interface is decreased. For the configuration in group 1, the atoms at the interfaces are randomly distributed, which forms a mixed interface. For the configuration in group 8, every two Ge atoms at the interfaces can be connected by the path forming only by the Ge atoms, like the atom A and B shown in Fig. 5(c), which is in an unmixed state. With the evolution of optimization, the interfaces gradually turn from mixed to unmixed state. Previous studies have reported that the acoustic mismatch between Si/Ge layers can impede the phonon transport [58]. The unmixed roughness can induce large acoustic mismatch, resulting in the reduction of thermal conductivity. On the contrary, the mixed roughness can decrease the acoustic mismatch of Si/Ge layers and enhance the thermal conductivity.

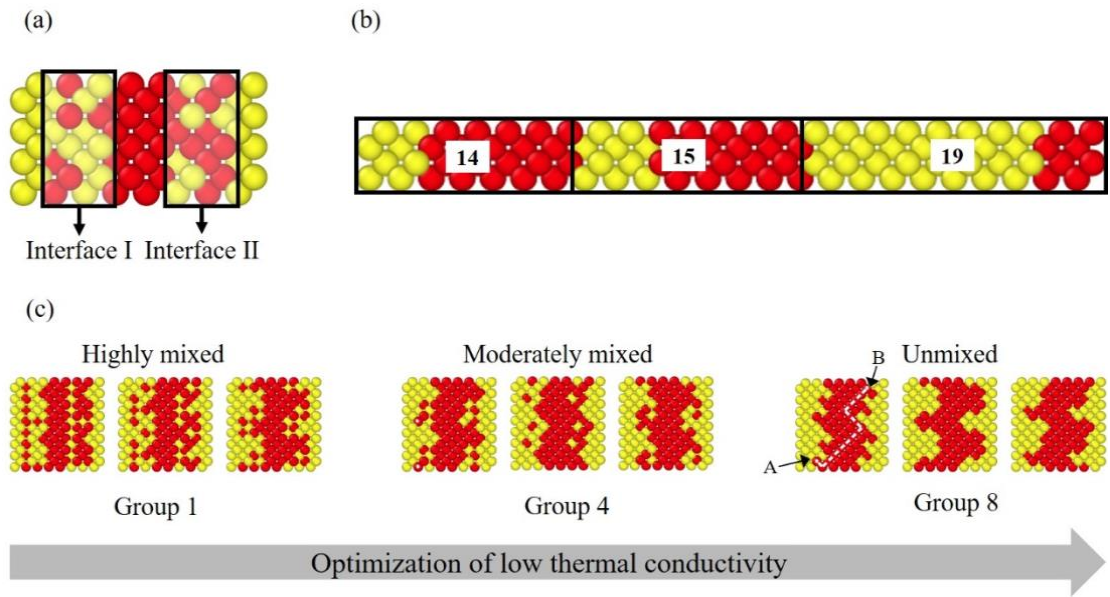

**Fig. 5** (a) The unit cell (4 × 2 × 1) for optimization of interface roughness. The black boxes show the interfaces. (b) The unit cell (12 × 1 × 1) for the optimization of layer thickness arrangement. A black box denotes a "layer", which is composed of multiple layers of Si and Ge. The layer thickness refers to the number of continuous monolayers of Si and Ge. From left to right, the configuration has three layers with thicknesses of 14, 15 and 19 and thus the average layer thickness is 16. (c) Three sets of representative configurations, which are from the group 1, 4 and 8. In group 1, the atoms at the interfaces are mixed, while in group 8, the interfaces are in an unmixed state. For unmixed roughness, every two Ge atoms at the interfaces, like the atom A and B, can be connected by the path forming only by the Ge atoms. With the evolution of optimization, the interfaces gradually turn from a mixed state to an unmixed state.

Next, the optimization of the arrangement of alternating layer thickness is performed in the alloy systems with 12 × 1 × 1 unit cells. Twelve unit cells along the *x*-direction are set to provide a large design degree of freedom for optimization. For simplicity, we only consider the smooth interface, and the schematic of the optimization system is shown in Fig. 5 (b). Note that here the "layer" refers to a unit composed of multiple layers of Si and Ge, which are denoted by the black boxes in Fig. 5 (b). The layer thickness refers to the number of continuous monolayers of Si and Ge. The system can be characterized by the average layer thickness $\bar{l}$. For example, as shown in Fig. 5(b), the configuration has three layers with thicknesses of 14, 15 and 19 and thus $\bar{l}$ of the system is 16.

We perform optimization for the systems with $\bar{l}$ of 12, 16, and 24. For each system, the different configurations are generated by randomly assigning the monolayers as Si or Ge. In order to reduce the degree of freedom, we limit the minimum layer thickness as four. The optimization results are similar for these three systems. For simplicity, we only present the results of the system with $\bar{l}$ of 24. The optimal configuration with smooth interfaces and optimal layer thickness arrangement (SIOL structure) and corresponding thermal conductivity are presented in Fig. 6(b). As a comparison, the layered structure with smooth interfaces and equal layer thickness (superlattice) is shown in Fig. 6(a). The thermal conductivity of the SIOL structure is lower than that of the superlattice, which is consistent with previous studies [27,61]. The disorder arrangement of layer thickness can induce destructive interferences, which effectively weakens the heat conduction [61].

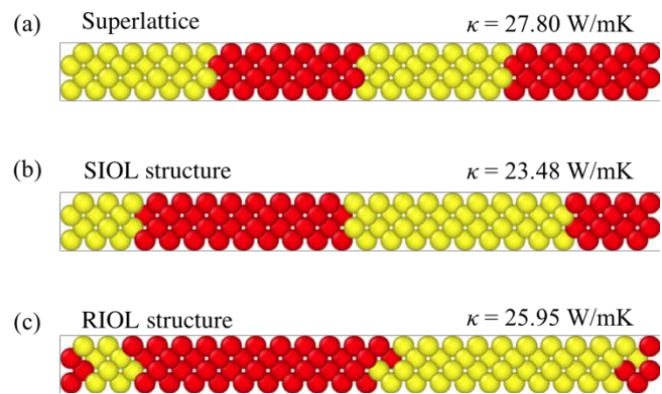

**Fig. 6** (a) The layered structure with smooth interfaces and equal layer thickness (superlattice). (b) The layered structure with smooth interfaces and optimal layer thickness arrangement (SIOL structure). (c) The layered structure with rough interfaces and optimal layer thickness arrangement (RIOL structure). The average layer thickness of these configurations is 24. The corresponding thermal conductivities are shown above the configurations.

**4.2 Coupling effect of interface roughness and layer thickness arrangement**

Based on the optimization of interface roughness and layer thickness arrangement separately, it is found that the layered structure with unmixed interface roughness and particular layer thickness arrangement can have low thermal conductivity. We further identify the specific atomic configuration with the lowest thermal conductivity considering both two factors.

We first optimize the interface roughness based on obtained SIOL structure. The SIOL structures with $\bar{l}$ of 16 and 24 are considered. In order to generate candidate configurations, the atoms in the outer monolayers are randomly assigned as Si or Ge while maintaining the composition. With Bayesian optimization, it is found the thermal conductivity is not further reduced and the optimal configuration still is the original SIOL structure.

We compare the detailed atomic configurations of the calculated candidates and the original SIOL structure. The deviations are quantified by the number of atoms that are different at the same lattice sites. The deviations for $\bar{l}$ being 16 and 24 are shown in Fig. 7 (a) and (b), respectively. For both scenarios, overall, the configurations with higher thermal conductivities possess larger deviations from the original SIOL structures. As mentioned early, the particular layer thickness arrangement in SIOL structure can impede the thermal transport by the destructive phonon interference. Adding interface roughness may destroy the destructive phonon interference, leading to the enhancement of thermal conductivity[61]. On the other hand, adding interface roughness can increase the phonon-interface scattering, inducing the reduction of thermal conductivity[60]. From the optimization, it seems that the interface roughness mainly interrupts the phonon destructive interference rather than increases the incoherent phonon scattering.

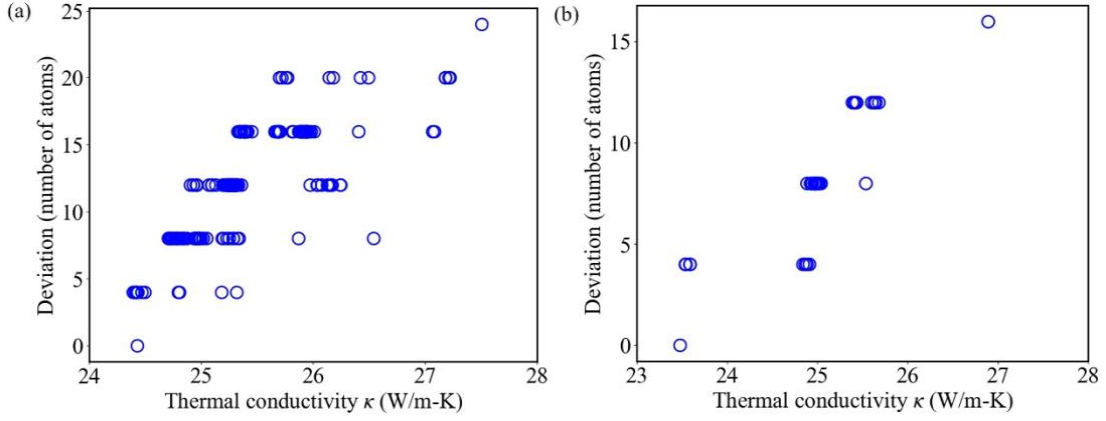

**Fig. 7** The deviations between the original SIOL structure and calculated candidate structures in Bayesian optimization with $\bar{l}$ being 16 (a) and 24 (b).

Similarly, we perform the optimization of layer thickness arrangement based on the configuration with unmixed interfaces. For generality, the shapes of roughness are randomly produced. The configurations of the 12 × 1 × 1 unit cell with $\bar{l}$ being 16 and 24 are considered. With Bayesian optimization, the configuration with rough interfaces (unmixed) and optimal layer thickness arrangement (RIOL structures) is found, which is shown in Fig. 6 (c). Compared with the SIOL structure, the thermal conductivity of RIOL structure is higher, which confirms that adding rough interfaces are not beneficial for reducing the thermal conductivity. The optimization results are the same for the cases with $\bar{l}$ being 16. With the comprehensive search, the atomic configuration with the lowest thermal conductivity is identified as the layered structure with smooth interfaces and optimal layer thickness arrangement, namely SIOL structure.

## 5. Conclusions

In this study, we search for the lowest thermal conductivity atomic configuration of $Si_{0.5}Ge_{0.5}$ alloy system with first-principles calculations and Bayesian optimization. The optimizations of the atomic configuration of different unit cell sizes are carried out and the layered structures are found to possess lower thermal conductivities. The flattening of the phonon dispersion curve and the filtering effect of alternate Si/Ge layers of the layered structure are responsible for the low thermal conductivity. Furthermore, the layer thickness arrangement and interface roughness of the layered structure are optimized. Through a comprehensive search, we unambiguously proved that the optimal configuration of the alloy system with the lowest thermal conductivity is the layered structure with smooth interfaces and optimal layer thickness arrangement. This study also demonstrates the effectiveness of Bayesian optimization for material design at the atomic scale.

**Conflict of interest**

There are no conflicts to declare.

**Supporting Information**

Supplementary data associated with this article can be found in the online version at XXX.

**Acknowledgements**

This work is supported by the National Natural Science Foundation of China (No. 51676121), Guangdong Province Key Area R&D Program (2019B010940001) and the Materials Genome Initiative Center of Shanghai Jiao Tong University. Simulations were performed with computing resources granted by HPC ($\pi$) from Shanghai Jiao Tong University.

# Supporting Information: Seeking for low thermal conductivity atomic configurations in $Si_{0.5}Ge_{0.5}$ alloys with Bayesian Optimization

**Jiahao Yan,[a#] Han Wei,[a#] Han Xie,[a] Xiaokun Gu [b] and Hua Bao [a*]**

[a] *University of Michigan-Shanghai Jiao Tong University Joint Institute, Shanghai Jiao Tong University, Shanghai 200240, China.*

[b] *Institute of Engineering Thermophysics, School of Mechanical Engineering, Shanghai Jiao Tong University, Shanghai 200240, China*

---

\* Corresponding Author

E-mail: hua.bao@sjtu.edu.cn. #These authors contributed equally.

## 1. Bayesian optimization

Bayesian optimization is designed for the "black-box" global optimization. The strategy in Bayesian optimization is to build a surrogate model for the unknown objective function. Generally, Gaussian process regression is applied to build the surrogate to model the mapping from descriptors to the function evaluations (i.e. thermal conductivities in our work) of the existing candidates. Then, an acquisition function constructed from the surrogate model is used to decide the next candidates to sample, which are further added to the existing dataset to update the surrogate model[1,2]. By repeating this procedure, the optimal solution (i.e. optimal atomic configurations with the lowest thermal conductivity in our work) could be quickly obtained.

In this work, 1D arrays are used to characterize the atomic configurations as the descriptors, which consist of "1" and "2" representing Si and Ge atoms/layers, respectively. The surrogate model that maps from the descriptors to the lattice thermal conductivities of different configurations is established by Gaussian process regression. We use expected improvement[2] as the acquisition function to decide the optimized candidate configurations.

## 2. Methods to calculate the lattice thermal conductivity

In Bayesian optimization-based framework, the lattice thermal conductivities of different configurations are obtained by the combination of the Boltzmann transport equation (BTE) and the Fourier's law[3]

$$k_{\alpha\beta} = \sum_\lambda c_\lambda v_{\lambda,\alpha} v_{\lambda,\beta} \tau_\lambda, \qquad (1)$$

where $\lambda$ denotes the phonon mode marked by phonon branch $v$ and wave vector $\boldsymbol{q}$. $c_\lambda$ is the volumetric heat capacity. $v_{\lambda,\alpha}$ and $v_{\lambda,\beta}$ are the phonon group velocity vector in $\alpha$ and $\beta$ directions, respectively. $\tau_\lambda$ is the phonon relaxation time. $c_\lambda$ and $v_\lambda$ only require harmonic interatomic force constants (IFCs)[4]. $\tau_\lambda$ needs both harmonic and anharmonic IFCs, which is depended on the size of different unit cells. Note that the difference of interatomic force constants (IFCs) of Si and Ge are minor comparing to the mass difference. For simplicity, the IFCs for different atomic configurations are set as the same as Si unit cells with corresponding sizes and only the mass difference is considered. The IFCs for large unit cells are carefully mapped from the Si IFCs of the

FCC unit cell. For 2×1×1 unit cell, $\tau_\lambda$ is calculated by considering three-phonon scattering processes under single mode relaxation time approximation (SMRTA)[5]. The expression is given by[6]

$$\frac{1}{\tau_\lambda} = \sum\nolimits_{\lambda'\lambda''}^{+} \frac{2\pi}{\hbar^2}(n_{\lambda'}^0 - n_{\lambda''}^0) \times |V_{\lambda,\lambda'\lambda''}^{+}|^2 \delta_{q_\lambda+q_{\lambda'}-q_{\lambda''},\mathbf{G}} \, \delta(\omega_\lambda + \omega_{\lambda'} + \omega_{\lambda''}) + \sum\nolimits_{\lambda'\lambda''}^{-} \frac{\pi}{\hbar^2}(1 + n_{\lambda'}^0 + n_{\lambda''}^0) \times |V_{\lambda,\lambda'\lambda''}^{-}|^2 \delta_{q_\lambda-q_{\lambda'}-q_{\lambda''},\mathbf{G}} \, \delta(\omega_\lambda - \omega_{\lambda'} - \omega_{\lambda''}), \qquad (2)$$

where $\hbar$ denotes the reduced Plank constant. $n_\lambda^0$ denotes the equilibrium phonon distribution. $\mathbf{G}$ is a reciprocal lattice vector, and $V_{\lambda,\lambda'\lambda''}^{\pm}$ are the three-phonon coupling matrixes. As shown by equation (2), the anharmonic scattering process is classified into two parts. The first summation represents the process that phonon mode $\mathbf{q}_\lambda$ decays into phonon mode $\mathbf{q}_{\lambda'}$ and $\mathbf{q}_{\lambda''}$. The second summation represents the process that phonon mode $\mathbf{q}_\lambda$ absorbs phonon mode $\mathbf{q}_{\lambda'}$ then yields $\mathbf{q}_{\lambda''}$. In each part, the first $\delta$ is the Knonecker delta function, and the second $\delta$ is the Dirac delta function. The former is used to denote the normal processes and umklapp processes through $\mathbf{G} = 0$ and $\mathbf{G} \neq 0$, respectively, and the latter represents the conservation of energy.

In this work, the harmonic and anharmonic interatomic force constants (IFCs) are obtained by density-functional perturbation theory (DFPT) with 2×2×2 supercell and 8×8×8 k-grid. The calculations are carried out by phonopy[7] and VASP package[8] with GGA pseudopotentials[9], and two-nearest neighbors were considered in the calculation of anharmonic IFCs.

Note that the computational costs of thermal conductivities are unaffordable for unit cells larger than 2×1×1 unit cells, thus the calculation is simplified by employing a simplified phonon relaxation time model proposed by Klemens. The expression of the new relaxation time model is given by[10]

$$\tau_\lambda^{-1} = B\omega^2 T^3 e^{-\frac{\theta}{\alpha T}}, \qquad (3)$$

where $B$ is a fitting constant, $\omega$ denotes phonon frequency, $T$ denotes temperature, $\theta$ denotes Debye temperature[11] and $\alpha$ is a numerical constant. Parameter $B$ and $\alpha$ are fitted by the least-squares method with a set of accurate data from full first principle calculation, while the frequencies and the Debye temperature are carefully calculated to ensure the accuracy is good enough to distinguish different structures. The Debye temperature is in the form of expression proposed by Domb and Salter[11]

$$\theta = \sqrt{\frac{5}{3}\frac{\hbar^2}{k_B}\frac{\int_0^\infty \omega_{\lambda,A}^2 D(\omega_{\lambda,A})d\omega}{\int_0^\infty D(\omega_{\lambda,A})d\omega}}, \qquad (4)$$

where $\hbar$ denotes the reduced Plank constant, $k_B$ is Boltzmann constant, and $D(\omega_{\lambda,A})$ denotes the density of state for acoustic modes.

To validate the accuracy of this model, we calculate the room temperature (300 K) thermal conductivities of all candidates in the 2×1×1 unit cell with both full first principle calculation and the simplified model. The results are shown in Fig. S1. The thermal conductivities calculated by the simplified model are close to that calculated by first principle calculation. Spearman correlation coefficient[12] is used to measure the correlation between the results from the simplified model and full calculation, of which the value is beyond 0.9, showing a strong correlation. In our work, the difference of rank of thermal conductivity is more important than the absolute values. We consider the simplified model as a viable way to calculate the thermal conductivities, where the sacrifice of accuracy within a controlled range is acceptable. Therefore, we apply this method for calculating the thermal conductivities of configurations in unit cells larger than 2×1×1.

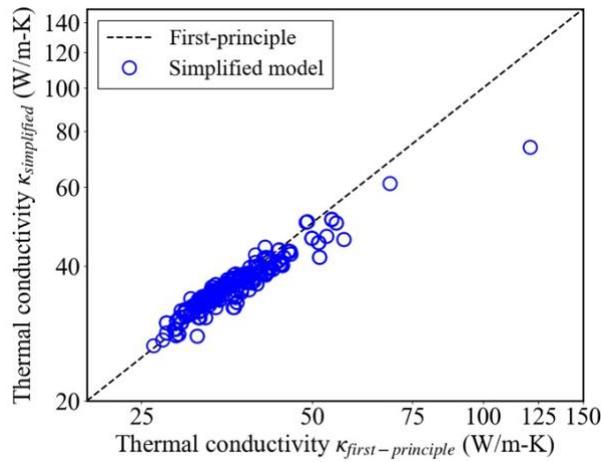

**Fig. S1** The comparison of thermal conductivities calculated by first principle calculation approach and the simplified model at 300 K.

**3. Group neighboring atoms for the reduction of candidates**

The number of candidates increases rapidly for the optimizations of systems in 2×2×1 (~ $10^9$) and 2×2×2 unit cells (~ $10^{18}$), thus we group neighboring atoms as new units to reduce the total number of candidates. Fig. S2 shows the grouping schemes of atoms for 2×2×1 and 2×2×2 unit cells, where the black boxes denote the new units. In order to keep the space of structural adjustment in the direction of thermal conductivity, the black boxes for grouping are thin in that direction. For 2×2×1 unit cell, every two neighboring atoms are grouped so that 16 groups are generated, including 8 Si groups and 8 Ge groups. Similarly, for 2×2×1 unit cell, every four neighboring atoms are grouped, yielding 8 Si groups and 8 Ge groups. The number of candidates is able to be reduced to 12870 by replacing atoms with groups, which corresponds to an affordable computational cost.

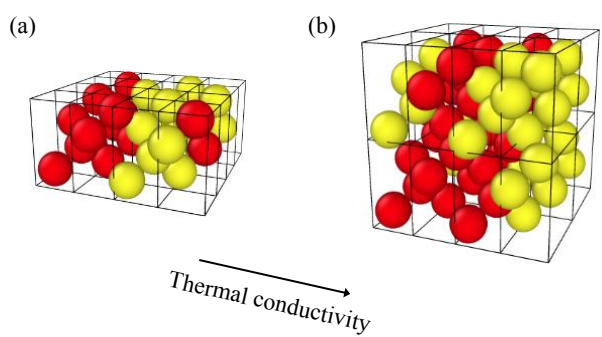

**Fig. S2** The schematic diagram of grouping schemes for (a) 2×2×1 and (b) 2×2×2 unit cells, where the black boxes denote the new units for manipulation.

**Supplementary References**